%% file: main_v3.tex
\documentclass[reprint, twocolumn, amsfonts, amsmath, amssymb, superscriptaddress, aps]{revtex4-2}

\usepackage{graphicx}% Include figure files
\usepackage{dcolumn}%  Align table columns on decimal point
\usepackage{bm}% bold math
\usepackage[normalem]{ulem}
\usepackage{mathtools}
\usepackage[dvipsnames]{xcolor}
\usepackage[linkcolor=Blue,citecolor=Blue,urlcolor=Blue,colorlinks=true]{hyperref}
\usepackage[normalem]{ulem}
\usepackage{mathrsfs}

\newif\ifshownotes
\ifshownotes
	\newcommand{\note}[3]{{\color{#2}[#1: #3]}}
    
    \newcommand{\del}[3]{\textbf{\color{#2}\sout{#3}}}
    \newcommand{\eqdel}[3]{\textbf{\color{#2}#3}}
\else
	\newcommand{\note}[3]{}
	\newcommand{\del}[3]{}
	\newcommand{\eqdel}[3]{}
    
\fi

\usepackage{xr}

\begin{document}
\title{Coherent advantage in the computational expressivity of excitonic networks}
\author{Matthew Du}
\email{madu@uchicago.edu}
\affiliation{Department of Chemistry, University of Chicago, Chicago, Illinois 60637, USA}
\affiliation{The James Franck Institute, University of Chicago, Chicago, Illinois 60637, USA}
\author{Carlos Floyd}
\affiliation{Department of Chemistry, University of Chicago, Chicago, Illinois 60637, USA}
\affiliation{The James Franck Institute, University of Chicago, Chicago, Illinois 60637, USA}
\author{Dipti Jasrasaria}
\affiliation{Department of Chemistry, University of Chicago, Chicago, Illinois 60637, USA}
\affiliation{The James Franck Institute, University of Chicago, Chicago, Illinois 60637, USA}
\author{Suriyanarayanan Vaikuntanathan}
\email{svaikunt@uchicago.edu}
\affiliation{Department of Chemistry, University of Chicago, Chicago, Illinois 60637, USA}
\affiliation{The James Franck Institute, University of Chicago, Chicago, Illinois 60637, USA}
\date{\today}
\begin{abstract}
The rising energy consumption of AI has generated interest in physical systems as alternative substrates for trainable computation. 
Recent experimental advances have enabled precise control over the couplings between molecular chromophores, which give rise to coherent excitation dynamics. 
Here, we study driven-dissipative excitonic networks as a computational platform, where the intersite couplings define the input and the steady state defines the output. 
We show that coherence enables computational expressivity to scale with network size, analogous to artificial neural networks. 
Both this scaling and the overall expressivity are suppressed by strong dephasing.
Our work establishes coherence as a resource for expressive computation in nonequilibrium quantum systems.
\end{abstract}
\maketitle
\section{Introduction}
\begin{figure}
\centering\includegraphics{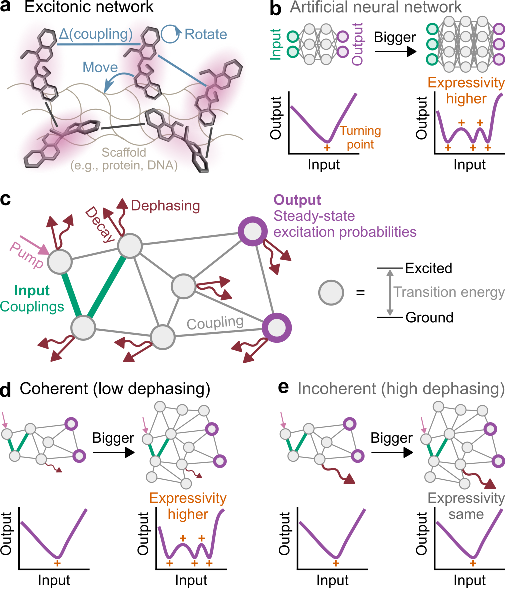}
\caption{
\textbf{Overview: computational expressivity of open quantum networks.}
(a) In an excitonic network of molecular chromophores, excitations become delocalized through coherent (wave-like) transport mediated by dipole-dipole coupling.
Using a scaffold (e.g., protein or DNA), the couplings can be tuned by adjusting the position or orientations of the chromophores. 
(b) The computational expressivity of an artificial neural network increases with model size. 
Here the expressivity is measured by the number of turning points. 
(c) Excitonic network as a computational platform.
Each site is associated with a transition energy of excitation and couplings to other sites. 
Excitations are pumped into the system at a subset of sites but can decay and dephase at all sites due to interaction with their local environment.
The input of the computation is a subset of couplings, and the output is a subset of site excitation probabilities at steady state.
(d, e) Effect of system size on the computational expressivity.
The expressivity increases with system size in the coherent regime of low dephasing (d), mirroring artificial neural networks, but is invariant to system size in the incoherent regime of high dephasing (e). 
\label{fig:expressivity-schematic}
}
\end{figure}

The ever-increasing size of artificial intelligence (AI) models is driven by a simple principle: bigger models perform better. 
At their core, AI models rely on artificial neural networks, nonlinear functions capable of exhibiting increasingly complex behavior as their parameter count grows (Fig.~\ref{fig:expressivity-schematic}b)~\cite{goodfellow_deep_2016}. 
However, this improved expressivity comes at the expense of greater energy consumption.

In a subfield of neuromorphic computing~\cite{markovic_physics_2020}, physical analogs of neural networks are being explored as energy-efficient alternatives to their digital counterparts~\cite{momeni_training_2025}.
Typically, a physical substrate is trained such that its response to (or, equivalently, transformation of) different values of an external stimulus approximates the same target function as a conventional neural network would~\cite{momeni_training_2025}. 
Other approaches involve splitting either the external stimulus~\cite{wright_deep_2022} or the internal parameters~\cite{wanjura_fully_2024,yildirim_nonlinear_2024,xia_nonlinear_2024,floyd_limits_2025} into two subsets, training one to produce the desired response to the other.
In all of these schemes, the response requires energy to sustain but is ultimately shaped by the intrinsic dynamics of the physical system, potentially yielding significant energy savings~\cite{de-lima_machine_2019,dubcek_in-sensor_2024}.  
Physical neural networks have been realized in a plethora of substrates, including electronic~\cite{sebastian_memory_2020}, optical~\cite{de-lima_machine_2019,wetzstein_inference_2020}, spintronic~\cite{sengupta_neuromorphic_2018}, mechanical~\cite{wright_deep_2022,dubcek_in-sensor_2024}, and chemical~\cite{okumura_nonlinear_2022,cherry_supervised_2025} systems.   

Here we explore the potential of excitonic networks for neuromorphic computing, motivated by recent experimental advances that now make it possible to train these open quantum systems. 
In excitonic networks, exemplified by photosynthetic light-harvesting complexes~\cite{engel_evidence_2007,mirkovic_light_2016,jang_delocalized_2018}, chromophores coupled by dipole-dipole interactions support coherent (wave-like) excitation transport (Fig.~\ref{fig:expressivity-schematic}a), while environmental dephasing drives a crossover toward incoherent (dissipative) dynamics that, in the strong-dephasing limit, reduce to a classical rate model~\cite{cao_optimization_2009,hoyer_limits_2010}.
Over the last decade, researchers have advanced the use of DNA scaffolds as synthetic platforms on which chromophores can be precisely arranged into specified positions and orientations~\cite{mathur_pursuing_2023,hart_engineering_2023} (Fig.~\ref{fig:expressivity-schematic}a).
These parameters, in turn, set the coupling strength between chromophores and, thereby, the dynamical regime. 
This tunability has been used to design discrete logic gates for digital quantum computing~\cite{castellanos_design_2020,yurke_implementation_2023}.
We instead consider how it could be exploited for neuromorphic computing.
Programmable interactions enable the excitation dynamics to be trained on a task while also allowing us to isolate how coherence shapes expressivity.

Here we show that coherence can serve as a physical resource for expressive computation in open quantum systems.
We study driven-dissipative excitonic networks as a computational platform in which intersite couplings serve as inputs and steady-state excitation probabilities serve as outputs (Fig.~\ref{fig:expressivity-schematic}c).
In the coherent regime, the computational expressivity grows with network size, mirroring the scaling seen in digital neural networks (Fig.~\ref{fig:expressivity-schematic}b,d).
In the incoherent regime, by contrast, expressivity becomes independent of system size, recovering the behavior of classical networks at steady state~\cite{wanjura_fully_2024,floyd_limits_2025} (Fig.~\ref{fig:expressivity-schematic}e).
Our results identify a concrete mechanism by which the dynamical regime of a physical system controls its expressivity.

\section{Computational expressivity grows with system size}
\label{sec:quantum-expressivity}
An assembly of molecular chromophores under continuous excitation evolves toward a steady state that depends on chromophore energies, interchromophoric coupling, and dissipative processes.
We will characterize the response of the steady state to varying coupling between chromophores, elucidating the breadth of functions that this input-output relation can represent for computation.  
Encoding the input in system parameters such as coupling strengths rather than the excitation source is expected to yield greater nonlinearity~\cite{wanjura_fully_2024,yildirim_nonlinear_2024,xia_nonlinear_2024}, bypassing limitations such as the generally weak optical nonlinearity of molecules.
As discussed above, the coupling between chromophores is tunable through their relative positions and orientations~\cite{mathur_pursuing_2023,hart_engineering_2023}.
We take inspiration from studies on the interplay between coherent and incoherent dynamics in light-harvesting systems~\cite{plenio_dephasing_2008,mohseni_environment_2008,cao_optimization_2009,manzano_quantum_2012}, asking not how such networks transport excitations but how they can use them to compute.

We consider networks of $N$ molecular chromophores whose coherent dynamics are governed by the Hamiltonian 
\begin{equation}
    H = \sum_{j=1}^N E_j \sigma_j^\dagger \sigma_j + \sum_{j=1}^N \sum_{l> j} V_{jl} \left(\sigma_j^\dagger \sigma_l + \sigma_l^\dagger \sigma_j\right),
    \label{eq:h}
\end{equation}
where $\sigma_j^\dagger$ and $\sigma_j$ are the creation and annihilation operators for an electronic excitation on site $j$, respectively; $E_j$ is the excitation energy of site $j$; and $V_{jl}$ is the (real-valued) coupling for excitation hopping between sites $j$ and $l$.
The couplings create coherences between different sites, driving localized excitations into delocalized superpositions. 
In addition to the coherent dynamics, excitations are pumped into just one site $j=1$ but can decay at all sites. 
We assume that pumping occurs at a rate much slower than decay, which is representative of ambient light and modest laser intensities.
Thus, we restrict the dynamics to the zero- and single-excitation manifolds, defined by $N+1$ basis states: $|0\rangle$, the ground state with no excitations, and $|j\rangle = \sigma_j^\dagger|0\rangle$, the state with only site $j$ excited, for $j=1,\dots,N$.
Lastly, the coupling between an electronic excitation and its local vibrational environment induces dephasing, the suppression of coherences without creating or destroying an excitation.
In molecules at ambient temperature, dephasing is the main source of decoherence, with excitation decay a minor contributor.
Because the interchromophoric couplings sustain coherences while dephasing suppresses them, the two compete, generally leading to complex dynamics involving the delocalized eigenstates of $H$.

\begin{figure*}
\centering\includegraphics{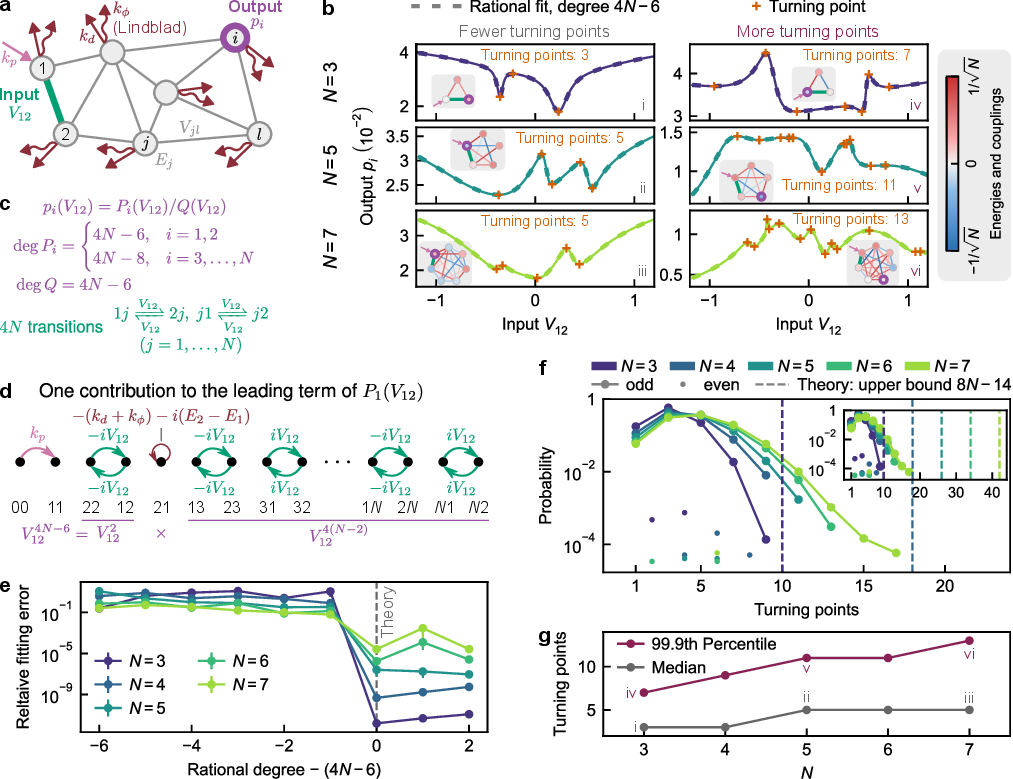}
\caption{\textbf{Computational expressivity grows with system size in the coherent regime.} 
(a) Model of excitonic networks as a computational platform.
An excitation on site $j=1,\dots,N$ has transition energy $E_j$ and can hop to site $l$ according to coupling $V_{jl}=V_{lj}^*$.
Excitations are pumped into site 1 at rate $k_p$ and decay from any site at rate $k_d$.
Excitations at each site also dephase at a constant rate $k_\phi$, as described by the Lindblad master equation.
The input of the computation is the coupling $V_{12}$ between the pumped site 1 and a not-pumped site 2, and the output is the excitation probability $p_i$ of a site $i$ at steady state. 
(b) Examples of computational input-output relations $p_i(V_{12})$ for excitonic networks of different system sizes $N$.
The dashed lines are a fit to a rational function of degree $4N-6$, i.e., whose numerator and denominator are polynomials of degree $4N-6$.
The orange plus signs indicate turning points. 
For each $N$, two representative networks with relatively few (left column) and more (right column) turning points are shown.
(c) The input-output relation can be expressed as a rational function $p_i(V_{12})=P_i(V_{12})/Q(V_{12})$, where  $P_i$ and $Q$ are polynomials with degrees $\sim 4N$. 
The $4N$ scaling reflects how there are $4N$ transitions mediated by the coupling $V_{12}$ or its complex conjugate $V_{21}$.
(d) Graph representation of one contribution to the leading term of $P_1$.
The contributions arise when expressing $P_1$ using the Leibniz formula for determinants (Appendix~\ref{app:rational}).   
(e) Relative error of fitting a rational function of varying degrees to input-output relations of random excitonic networks.
The error bars represent standard error of the mean.
(f, g) Histogram (f), 99th percentile (g), and median (g) of turning points in the input-output relations of random excitonic networks for varying system sizes $N$. 
In (f), the odd and even number of turning points are distinguished by markers with and without lines, respectively (see main text); and the dashed lines indicate the upper bound $8N-14$ of turning points as determined by the rational form of $p_i(V_{12})$.
In (g), the Roman numerals refer to the representative input-output relations shown in (b) for different numbers of turning points.
\label{fig:quantum-expressivity}}
\end{figure*}

To enable insights from analytical theory, we work primarily in a regime which includes high temperature where the dephasing, and thus the full excitation dynamics, can be described by a Lindblad master equation~\cite{manzano_quantum_2012} (Fig.~\ref{fig:quantum-expressivity}a; Appendix~\ref{app:lindblad-from-redfield}),
\begin{align}
    \dot{\rho}&=\mathcal{L}\rho \nonumber\\
    &=-i[H,\rho]
    +k_p\mathcal{D}[\sigma_1^\dagger]\rho
    +\sum_{j=1}^N \left(k_d\mathcal{D}[\sigma_j]+k_\phi\mathcal{D}[\sigma_j^\dagger \sigma_j]\right)\rho,
    \label{eq:lindblad}
\end{align}
with the convention $\hbar=1$; below, we explore a more general regime by considering a Redfield description of dephasing~\cite{redfield_theory_1957,breuer_theory_2002}.
Here, $\mathcal{L}$ is the Liouvillian superoperator governing the full dynamics, $\rho$ is the density matrix representing the state of the system, and $\mathcal{D}[L]\rho=L\rho L^\dagger-\frac12\{L^\dagger L,\rho\}$ is a Lindblad superoperator defined by operator $L$.
The first term represents the coherent dynamics through $H$, while the remaining terms represent the incoherent processes.
Pumping, decay, and dephasing occur at rates $k_p$, $k_d$, and $k_\phi$, respectively; the latter two are taken to be uniform across all chromophores.
Since $H$ depends on the couplings $V_{jl}$, so does the Liouvillian $\mathcal{L}(\{V_{jl}\})$ and hence the steady state $\rho^{ss}(\{V_{jl}\})$.

We proceed to study the computational expressivity of the steady-state excitation probabilities, $\{p_i(\{V_{jl}\})\}\equiv \{\rho_{ii}^{ss}(\{V_{jl}\})\}$ for $i\geq 1$.
(Hereafter, $i$ in the subscript denotes the output site, not the imaginary unit.) 
% We consider the couplings to be freely tunable for now, addressing their geometric correlations, i.e., positions and orientations of chromophores, in Sec.~\ref{sec:proposal}.
For concreteness, we focus on the one-dimensional relation $p_i(V_{12})$, where $V_{12}$ couples the pumped site ($j=1$) and an unpumped site ($j=2$).
We have checked that our results remain qualitatively unchanged if the input is instead a coupling between two unpumped sites, e.g., $V_{23}$.

Our central analytical result is that the input-output relation is a rational function,
\begin{equation}
    p_i(V_{12})
    =
    \frac{P_i(V_{12})}
         {Q(V_{12})},
    \label{eq:rational-main}
\end{equation}
where $P_i$ and $Q$ are polynomials in $V_{12}$ whose degrees scale linearly with system size,
\begin{equation}
\deg P_i =
\begin{cases}
4N - 6, & i=1,2\\
4N - 8, & i=3,\ldots,N
\end{cases},
\qquad
\deg Q = 4N - 6,
\label{eq:degree-main}
\end{equation}
implying that the nonlinearity of the response grows linearly with $N$ (Fig.~\ref{fig:quantum-expressivity}c).
We derive these results in Appendix~\ref{app:rational}.

Intuitively, the linear scaling of the polynomial degrees reflects the number of transitions mediated by the input coupling $V_{12}$ between density matrix elements.
Since $V_{12}$ couples states $|1\rangle$ and $|2\rangle$ [Eq.~\eqref{eq:h}], it induces $4N$ transitions between the excited-state elements of the density matrix ($\rho_{jl}$): $1j\rightleftharpoons 2j$ and $j1\rightleftharpoons j2$ for each site $j=1,\dots,N$ (Fig.~\ref{fig:quantum-expressivity}c), which correspond to coherence transitions when $j=3,\dots,N$.
Formally, this scaling is related to the number of times the input coupling appears in the Liouvillian $\mathcal{L}$ [Eq.~\eqref{eq:lindblad}] and can be seen by expressing the steady state in terms of sums over Liouvillian pathways (Fig.~\ref{fig:quantum-expressivity}d, Appendix~\ref{app:rational}).
In contrast, in the systems of~\cite{floyd_limits_2025,wanjura_fully_2024}, the input modifies a system-size-independent number of dynamical matrix elements, so the degrees of their rational input-output relations do not scale with system size.

Other aspects of the rational form admit a physical interpretation too.  
The common denominator $Q$ [Eq.~\eqref{eq:rational-main}] is a normalization arising from conservation of probability, as observed in chemical reaction networks~\cite{floyd_limits_2025}. 
The constant terms ($-6$ and $-8$) of the degrees [Eq.~\eqref{eq:degree-main}] reflect the large-$|V_{12}|$ behavior determined by density matrix perturbation theory~\cite{li_perturbative_2014} (Appendix~\ref{app:large-v12}): all population localizes on sites 1 and 2 while being depleted from all other sites as $V_{12}^{-2}$.

We numerically verify the rational form and the system-size scaling of its degree given by Eqs.~\eqref{eq:rational-main}-\eqref{eq:degree-main}.
We compute the input-output relation $p_i(V_{12})$ for random excitonic networks whose site energies and couplings are sampled independently from uniform distributions (Appendices~\ref{subapp:random-networks} and ~\ref{subapp:steady-state}). 
The energy differences, couplings, and dissipation rates are all scaled by $1/\sqrt{N}$ to keep the excited-eigenstate bandwidth and the relative timescales of coherent and dissipative dynamics constant with system size.
We work in the coherent regime, where the dephasing and decay rates are much smaller than the maximum absolute coupling of the distribution, $k_\phi, k_d \ll V_{\max} \equiv \max_{j,l} |V_{jl}|$.
We fit the input-output relations to rational functions whose numerator and denominator have a common degree (Appendix~\ref{subapp:fitting}), as shown in Fig.~\ref{fig:quantum-expressivity}b for networks of different sizes $N$. 
Numerically exact agreement is obtained at degree $4N-6$. 
As shown in Fig.~\ref{fig:quantum-expressivity}e, lower-degree rational functions fail to reproduce the input-output relation, consistent with the degree predicted by Eq.~\eqref{eq:degree-main}.

The growth of the rational function degree with system size suggests that the complexity of $p_i(V_{12})$ scales accordingly.
To quantify this complexity, we examine the number of turning points in the input-output relation (Fig.~\ref{fig:quantum-expressivity}b).
Turning points, where the slope switches sign, are a prototypical feature that computational models must capture to represent complex functions, such as energy surfaces. 
For ensembles of random networks, the turning-point count is predominantly odd (Appendix~\ref{app:parity}).
The entire distribution of turning points shifts towards larger values as $N$ increases (Fig.~\ref{fig:quantum-expressivity}f), with the strongest enhancement occurring in the high-expressivity tail of the distribution (Fig.~\ref{fig:quantum-expressivity}b,g).
This trend is corroborated by the upper bound $8N-14$ on the number of turning points, which is set by the degree of the derivative of $p_i(V_{12})$ and respected by the sampled counts (Fig.~\ref{fig:quantum-expressivity}f).
Thus, increasing the size of the excitonic network systematically increases the complexity of the input-output relations that can be realized, as in artificial neural networks (Fig.~\ref{fig:expressivity-schematic}b)~\cite{goodfellow_deep_2016}. 
\section{Coherent advantage in computational expressivity}
\label{sec:coherent-advantage}
The scaling of computational expressivity with system size observed for excitonic networks contrasts sharply with the behavior of chemical reaction networks studied by~\cite{floyd_limits_2025}, where the expressivity of the steady-state input-output relation does not increase with system size when the input and output are encoded similarly to here.
Since those reactions are described by a rate model with first-order kinetics, corresponding to the incoherent regime of excitonic networks~\cite{cao_optimization_2009,hoyer_limits_2010}, this discrepancy suggests that the scaling observed here originates from coherent dynamics.

\begin{figure*}
\centering\includegraphics{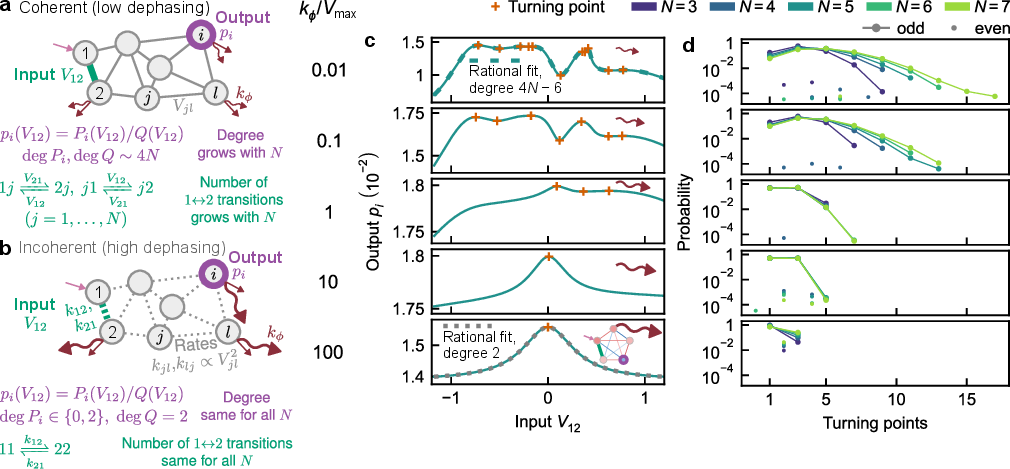}
\caption{\textbf{Coherent advantage in computational expressivity.}
(a, b) Computational expressivity of excitonic networks in the coherent regime of low dephasing (a) and the incoherent regime of high dephasing (b).
The input-output relation $p_i(V_{12})$ is given by a rational function whose degree and thus expressivity grows with system size $N$ in the coherent regime (a) but is independent of $N$ in the incoherent regime. 
The dependence on $N$ can be understood in terms of the number of $1\leftrightarrow 2$ transitions (i.e., mediated by $V_{12}$) in each regime.
(c) Examples of input-output relation for varying dephasing rate $k_\phi$ (in units of the maximum absolute coupling $V_{\max}$). 
The dashed lines are a fit to rational functions, and the orange plus signs indicate turning points. 
Increasing $k_\phi$ reduces computational expressivity as measured by the number of turning points (orange plus signs) and the degree of the rational form (dashed and dotted lines).
(d) Histogram of turning points in the input-output relations of random excitonic networks for varying $k_\phi$ and $N$. 
The odd and even number of turning points are distinguished by markers with and without lines, respectively (see main text).
\label{fig:coherent-advantage}
}
\end{figure*}

To investigate this connection, we examine how computational expressivity is affected by increasing the dephasing rate $k_\phi$, which promotes decoherence but not decay of excitations.
As $k_\phi$ increases, the number of turning points in the input-output relation $p_i (V_{12})$ decreases and eventually approaches one for all system sizes $N$ (Fig.~\ref{fig:coherent-advantage}c,d).
Thus, increasing dephasing not only suppresses computational expressivity but also eliminates its scaling with system size. 

We can analytically understand this suppression of expressivity by considering the limit of infinite dephasing, $k_\phi\rightarrow \infty$, in which coherences vanish, and the population dynamics reduce to a classical rate model~\cite{cao_optimization_2009,hoyer_limits_2010} (Appendix~\ref{subapp:dynamics-incoherent}),
\begin{equation}
    \dot{\rho}_{jj} = \sum_{l\neq j} k_{lj} \rho_{ll}-\left(\sum_{l\neq j}k_{jl}\right)\rho_{jj},
    \label{eq:kinetics}
\end{equation}
where, in addition to the pumping and decay rates $k_{0j}=\delta_{j1} k_p$ and $k_{j0}=k_d$, respectively, the transitions between excited states are described by the rates
\begin{equation}
    k_{jl}=k_{lj}=\frac{2(k_{d}+k_{\phi})V_{jl}^{2}}{(k_{d}+k_{\phi})^{2}+(E_{j}-E_{l})^{2}}, \quad j\neq0,\,l > j,
    \label{eq:k_jl}
\end{equation}
where only $k_{12}=k_{21}\sim V_{12}^2$ depends on the input coupling $V_{12}$.

Following a derivation analogous to Eqs.~\eqref{eq:rational-main}-\eqref{eq:degree-main} (Appendix~\ref{subapp:rational-incoherent}), we find that the input-output relation $p_i(V_{12})$ in the incoherent regime also takes the rational form of Eq.~\eqref{eq:rational-main}.
However, since excitations hop incoherently between chromophores at rates proportional to $V_{12}^2$ [Eq.~\eqref{eq:k_jl}], the numerator and denominator reduce to polynomials with degree 0 or 2 (Fig.~\ref{fig:coherent-advantage}b and bottom panel of Fig.~\ref{fig:coherent-advantage}c),
\begin{equation}
    \begin{split}
        P_i(V_{12}) &= 
        \begin{cases}
            m_i V_{12}^2+b_i, & i=1,2\\
            b_i, & i=3,\ldots,N
        \end{cases},\\
        Q(V_{12})&=m_Q V_{12}^2+b_Q,
    \end{split}
    \label{eq:p-incoherent}
\end{equation}
where $m_i$, $b_i$, $m_Q$, and $b_Q$ have the same sign (Appendix~\ref{subapp:rational-incoherent}). 
In particular, the input-output relation is symmetric about $V_{12}=0$ with a single turning point, regardless of $N$ (bottom panel of Fig.~\ref{fig:coherent-advantage}c). 
The computational expressivity in the incoherent regime is therefore not only lower than in the coherent regime but also independent of $N$. 

The stark difference between the two dynamical regimes can be understood in terms of the transitions mediated by the input coupling $V_{12}$ (Fig.~\ref{fig:coherent-advantage}a,b).
In the coherent regime, $V_{12}$ enters $4N$ Liouvillian transitions (Sec.~\ref{sec:quantum-expressivity}), with the linear scaling reflecting the growing number of coherences as the system size increases (Fig.~\ref{fig:coherent-advantage}a).
In the incoherent regime, by contrast, $V_{12}$ enters only 2 population transitions, between sites 1 and 2 (Fig.~\ref{fig:coherent-advantage}b).

So far, we have worked in a limit of high temperature (among other conditions).
Temperature, however, is a natural control knob for tuning dephasing: lowering it is a common strategy for reducing dephasing-induced broadening in spectroscopy.

\begin{figure}
\centering\includegraphics{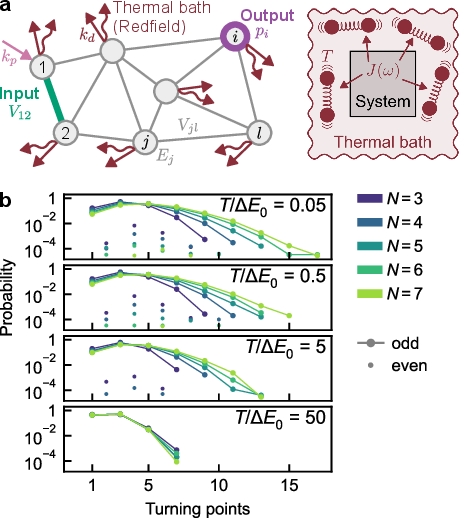}
\caption{\textbf{Coherent advantage in computational expressivity occurs at lower temperatures.}
(a) Model of excitonic network as a computational platform in the presence of dephasing induced by a thermal bath. 
The dephasing is characterized by temperature $T$ and spectral density $J(\omega)$, as described by the Redfield master equation. 
(b) Histogram of turning points in the input-output relations of random excitonic networks for varying $T$ (in units of $\Delta E_0 = 2V_\text{max}\sqrt{N}$, approximately the excited-eigenstate bandwidth; see Appendix~\ref{subapp:random-networks}) and system sizes $N$. 
The odd and even number of turning points are distinguished by markers with and without lines, respectively (see main text).
\label{fig:temperature-dependence}}
\end{figure}

To explore how temperature affects expressivity, we generalize the dephasing to the formalism of the (non-secular) Redfield master equation~\cite{redfield_theory_1957,breuer_theory_2002} (Appendix~\ref{app:redfield}), in which the dephasing rates depend on the transition energies of the excited eigenstates, temperature, and the spectral density~\cite{may_charge_2011}.  
The pumping and decay rates are kept unchanged, since we are interested in the effect of dephasing on expressivity. 
As temperature increases, the number of turning points in $p_i(V_{12})$ decreases and eventually ceases to increase with system size (Fig.~\ref{fig:temperature-dependence}b), mirroring the Lindblad results for increasing $k_\phi$. 

\section{Experimental Realization and Training}
\label{sec:proposal}
The computational framework developed above suggests a natural experimental realization in molecular excitonic aggregates, where excitonic couplings between chromophores can be controlled by tuning their relative positions and orientations~\cite{hart_engineering_2021}.
If one encodes the input into any one of these degrees of freedom, then it simultaneously modifies the couplings $V_{jl}$ between the displaced chromophore and all other sites in the network.
This automatically generates input multiplicity~\cite{wanjura_fully_2024,floyd_limits_2025}, i.e., the simultaneous modification of multiple Liouvillian elements by a single input, allowing additional scaling of expressivity beyond the system-size scaling established in Sec.~\ref{sec:quantum-expressivity}.
This stands in contrast to some previously studied platforms, where input multiplicity had to be introduced by explicitly replicating the input across independent degrees of freedom~\cite{wanjura_fully_2024,yildirim_nonlinear_2024,floyd_limits_2025}.

Training excitonic networks, i.e., optimizing parameters $\theta$ so that the input-output relation approximates a target function, can be carried out in various ways.
For computational optimization, an efficient approach is implicit differentiation of the steady state~\cite{vargas-hernandez_inverse_2020}: since the steady state satisfies the fixed-point condition $\mathcal{L}(\theta)\rho^{ss}(\theta)=0$ for all $\theta$, the gradient of any loss function with respect to $\theta$ can be obtained analytically, without requiring additional steady-state solves per parameter as finite-difference methods would.
For training an experimental system, computational optimization may miss aspects of the physical dynamics, though experimental measurements can be incorporated via methods such as backpropagation~\cite{wright_deep_2022}.
An alternative approach, which does not require knowing the exact equations of motion, is to employ contrastive update rules based on comparing steady states with and without an external perturbation that encodes the target function. 
Such rules have been applied to training various physical networks~\cite{stern_learning_2023} and recently extended to quantum systems~\cite{wanjura_quantum_2025,massar_equilibrium_2025,scellier_quantum_2026}.
\section{Conclusions}
\label{sec:conclusions}
In summary, we have studied networks of molecular chromophores under continuous excitation as a computational platform, where the input is encoded in the interchromophoric couplings and the output is encoded in the steady-state excitation probabilities.
These couplings, which are tunable through the relative positions and orientations of chromophores on platforms such as DNA scaffolds~\cite{mathur_pursuing_2023,hart_engineering_2023}, set both the dynamical regime and the complexity of the input-output relation.
We show that, in the coherent regime where the couplings sustain delocalized excitonic superpositions, the input-output relation is a rational function whose degree grows linearly with the number of chromophores.
Correspondingly, the number of turning points in the input-output relation increases with network size, indicating enhanced computational expressivity.
Electronic-vibrational coupling suppresses this scaling: as dephasing drives the system toward incoherent population transfer between chromophores, the expressivity reduces to a system-size-independent value described by a classical rate model. 
This scaling and its suppression persist under a Redfield description of dephasing that accounts for the temperature dependence of the vibronic coupling. 

These results establish coherence as a physical resource for the expressivity of steady-state computation.
In the incoherent limit, excitonic networks recover the behavior of classical networks~\cite{wanjura_fully_2024,floyd_limits_2025}, where expressivity does not scale with system size under similar input-output encodings and input multiplicity must be introduced by explicit replication.
Coherence intrinsically expands the number of dynamical pathways through which the input coupling influences the steady state as the network grows, so expressivity scales without any such replication. 
Yet coherence alone does not guarantee this scaling, since even coherent wave scattering systems require explicit input replication to achieve nonlinearity scaling with system size~\cite{wanjura_fully_2024,yildirim_nonlinear_2024}.
Understanding what distinguishes these cases, perhaps the choice of computational observable, is an interesting direction for future work. 

The computational platform proposed here naturally gives rise to a hybrid quantum-classical architecture.
The driven-dissipative excitonic network is a quantum substrate that provides an expressive forward map from couplings to steady-state populations, while a classical optimizer updates the physical parameters to approximate a target function (Sec.~\ref{sec:proposal}).
This division of labor parallels that of variational quantum circuits~\cite{cerezo_variational_2021,benedetti_parameterized_2019}, but with some key differences: the steady state of the driven-dissipative dynamics plays the role of the final state in a variational quantum circuit, and it arises from continuous evolution of the system rather than a sequence of discrete gate operations.
Excitonic networks also differ from quantum reservoir computing~\cite{mujal_opportunities_2021,sannia_dissipation_2024} and extreme learning machines~\cite{ghosh_quantum_2019} in that the physical parameters themselves are trainable, rather than fixed with only a classical readout layer being optimized.

Extending this framework beyond the single-excitation manifold is a natural future direction in the pursuit of quantum advantage.
The $M^\text{th}$ excitation manifold contains $\binom{N}{M}$ states, and $V_{12}$ connects $\binom{N-2}{M-1}$ pairs of states differing by a $1\leftrightarrow 2$ swap among $M$ occupied sites, yielding $\binom{N}{M}\binom{N-2}{M-1} = O(N^{2M-1})$ Liouvillian transitions (for $N\gg M$) and suggesting a polynomial growth of the computational expressivity with $N$ in contrast to the linear scaling found here.
This many-body extension would also move beyond the classical-wave correspondence of the single-excitation manifold~\cite{briggs_equivalence_2011},
opening the possibility of expressivity enhancements without classical analog.
More broadly, an open question is whether excitonic networks are universal approximators in the limit $N\rightarrow\infty$, given that the degree of the rational input-output relation grows without bound with system size and rational functions are known to approximate a broad class of functions~\cite{petrushev_rational_1987}.

\section*{Acknowledgments}
M.D. and S.V. were supported by DOE BES Grant No. DE-SC0019765. 
C.F. acknowledges support from the University of Chicago Data Science Institute AI + Science Research Initiative.
This work was completed using resources provided by the University of Chicago's Research Computing Center.

\appendix
\section{Dephasing in the formalism of the Redfield master equation}
\label{app:redfield}
Throughout this work, we primarily focus on the excitation dynamics given by Lindblad master equation~\eqref{eq:lindblad}, which is amenable to analytical treatment.  
However, the dephasing term in Eq.~\eqref{eq:lindblad}, 
\begin{equation}
    \mathcal{L}_\phi \equiv \sum_{j=1}^N k_\phi \mathcal{D}[\sigma_j^\dagger \sigma_j],
    \label{eq:lindblad-dephasing}
\end{equation}
is not valid across a broad range of parameters, particularly temperature.
To address this limitation, we consider a generalized dephasing term derived from the formalism of the Redfield master equation~\cite{redfield_theory_1957,breuer_theory_2002}.

Assuming that dephasing is induced by a site-local and site-independent harmonic bath coupled linearly to the system, the Hamiltonians governing the bath modes and their coupling to the system are
\begin{equation}
    H_B = \sum_{j=1}^{N}\sum_{k}\omega_{k}b_{jk}^{\dagger}b_{jk}
\end{equation}
and 
\begin{equation}
    H_{SB} = \sum_{j=1}^{N}\sum_{k}g_{k}\sigma_{j}^{\dagger}\sigma_{j}(b_{jk}^{\dagger}+b_{jk}),
\end{equation}
respectively, where $b_{jk}$ is the bosonic annihilation operator of the $k$th bath mode coupled to site $j$, $\omega_{k}$ is the mode frequency, and $g_{k}$ is the coupling strength. 
Following the standard Born (weak coupling) and Markov approximations and keeping on the real (dissipative) part of the one-sided Fourier transform of the bath correlation function, we can write the dephasing part of the Liouvillian governing the dynamics [$\mathcal{L}$, first line of Eq.~\eqref{eq:lindblad}] in the following form, which corresponds to the Redfield master equation~\cite{redfield_theory_1957,breuer_theory_2002}:
\begin{align} 
    \mathcal{R}_\phi = \sum_{j=1}^N  \sum_{\alpha\beta}\Gamma_\phi (E_\beta - E_\alpha)\hat{\mathcal{D}}_{\alpha\beta}[\sigma_j^\dagger \sigma_j].
    \label{eq:redfield-dephasing}
\end{align}
We note that this formulation does not rely on the secular approximation~\cite{breuer_theory_2002,may_charge_2011}.
The Redfield dephasing has a contribution from each transition $\alpha \rightarrow \beta$ between eigenstates $\alpha$ and $\beta$ of the system Hamiltonian $H$.
The temperature $T$ of the bath influences the dephasing through the transition-energy-dependent dephasing rate $\Gamma_\phi (E_\beta - E_\alpha)$, where
\begin{equation}
    \Gamma_\phi (\omega)=\Theta(\omega)J(\omega)[\bar{n}(\omega)+1]+\Theta(-\omega)J(-\omega)\bar{n}(-\omega),
    \label{eq:gamma-phi}
\end{equation}
$\Theta(\omega)$ is the Heaviside step function, 
\begin{equation}
    J(\omega)=\sum_k |g_{k}|^2 \delta(\omega-\omega_{k})
    \label{eq:j}
\end{equation}
is the spectral density governing the system-bath interaction~\cite{may_charge_2011}, and 
\begin{equation}
    \bar{n}(\omega)=\frac{1}{e^{\omega/T}-1}
    \label{eq:n-bar}
\end{equation}
is the average excitation number of the bath modes with frequency $\omega$, which corresponds to the Bose-Einstein distribution and $k_B = 1$.
Each dephasing contribution is also described by its own superoperator $\mathcal{D}_{\alpha \beta}$, which acts on the density matrix of the system according to
\begin{equation}
    \hat{\mathcal{D}}_{\alpha\beta}[L]\rho=(\hat{L}_{\alpha\beta}\rho L^\dagger - L^\dagger \hat{L}_{\alpha\beta}\rho)+\text{H.c.},
\end{equation}
where H.c. stands for Hermitian conjugate.
This transition-specific superopeartor $\mathcal{D}_{\alpha\beta}$ differs from the Lindblad superoperator $\mathcal{D}$ [defined below Eq.~\eqref{eq:lindblad}] in that it involves the eigenstate-projected jump operator $\hat{L}_{\alpha\beta}=|\alpha\rangle\langle\alpha|L|\beta\rangle \langle \beta|$. 

\section{Deriving Lindblad dephasing from Redfield dephasing}
\label{app:lindblad-from-redfield}
Here, we show that the Redfield dephasing superoperator [Eq.~\eqref{eq:redfield-dephasing}] reduces to the Lindblad dephasing superoperator [Eq.~\eqref{eq:lindblad-dephasing}], $\mathcal{R}_\phi = \mathcal{L}_\phi$, when the spectral density takes the Ohmic form~\cite{leggett_dynamics_1987} 
\begin{equation}
    J(\omega) =  \gamma\omega \exp(-\omega/\Omega_\text{cut})
    \label{eq:j-ohmic}
\end{equation}
for prefactor $\gamma$, and both its cutoff frequency $\Omega_\text{cut}$ and the temperature $T$ are much larger than the energy bandwidth $\Delta E=\max_\alpha E_\alpha - \min_\alpha E_\alpha$ of the excited eigenstates $\alpha$ of the system. 
For frequencies within the excited energy bandwidth, $|\omega| \leq \Delta E$, the two conditions imply that [Eq.~\eqref{eq:j}]
\begin{equation}
    J(\omega)\approx \gamma\omega,
\end{equation}
and  [Eq.~\eqref{eq:n-bar}]
\begin{equation}
    \bar{n}(\omega) \approx \frac{T}{\omega}\gg 1,
\end{equation}
respectively. 
Then the dephasing rate in Eq.~\eqref{eq:gamma-phi} becomes independent of transition energy, 
\begin{equation}
\Gamma_\phi(E_\beta - E_\alpha)=\gamma T,
\end{equation}
which is intuitive because the assumed conditions cause the frequency dependence of the spectral density and of the thermal excitation number to cancel each other out, while the high-temperature condition causes energetically uphill and downhill transitions between eigenstates to occur at the same rate. 
Thus, after comparing Eqs.~\eqref{eq:lindblad-dephasing} and~\eqref{eq:redfield-dephasing} and associating $k_\phi = 2\gamma T$, we obtain $\mathcal{R}_\phi = \mathcal{L}_\phi$.
\section{Dynamics under weak pumping}
\label{app:weak-pumping}
When the pumping rate is much smaller than the decay rate, $k_p\ll k_d$, we can approximately restrict the excitation dynamics to the zero- and single-excitation manifolds.
Within these manifolds and the Lindblad description of the dynamics [Eq.~\eqref{eq:lindblad}], the ground-state population ($\rho_{00}$), the excited-state populations ($\rho_{jj}$ for $j=1,\dots,N$), and the coherences between excited states ($\rho_{jl}=\rho_{lj}^*$ for $j,l=1,\dots,N$) evolve in one subspace according to
\begin{align}
\dot{\rho}_{00} & =k_{d}\sum_{j=1}^N\rho_{jj}-k_{p}\rho_{00},\label{eq:rho_00}\\
\dot{\rho}_{jj} & =\delta_{j1}k_{p}\rho_{00}-k_{d}\rho_{jj}-i\sum_{l\neq j}(V_{jl}\rho_{lj}-\rho_{jl}V_{lj}),\label{eq:rho_jj}\\
\dot{\rho}_{jl} & =-\left[(k_{d}+k_{\phi})+i(E_{j}-E_{l})\right]\rho_{jl}\nonumber\\
&\quad-i\left(\sum_{m\neq j}V_{jm}\rho_{ml}-\sum_{m\neq l}\rho_{jm}V_{ml}\right),\quad j\neq l,\label{eq:rho_jl}
\end{align}
while the coherences between the ground state and an excited state ($\rho_{j0}=\rho_{0j}^*$ for $j=1,\dots,N$) evolve in a separate subspace according to
\begin{align}
\dot{\rho}_{0j}=-\left[\frac{1}{2}(k_{p}+k_{d}+k_{\phi})-iE_{j}\right]\rho_{0j}+i\sum_{l\neq j}V_{lj}\rho_{0l}.
\end{align}
For symmetry, we have defined $V_{lj} \equiv V_{jl}$ for $j\neq0$ and $l > j$; we will continue to use this notation throughout the Appendix.
Since the ground-excited coherences decay but are not pumped, they vanish at steady state: $\rho_{0j}^{ss}=0$.

Throughout this work, we study the input-output relation in which the excited-state populations serve as outputs (see main text), and therefore focus largely on the dynamics of the corresponding reduced subspace, Eqs.~\eqref{eq:rho_00}-\eqref{eq:rho_jl}. 
We hereafter refer to the associated reduced Liouvillian as $\tilde{\mathcal{L}}$.
\section{Total excitation probability and ground-state population at steady state}
\label{app:tot-exc-00}
Here, we evaluate the total excitation probability and the ground-state population at steady state, which we will make use of in subsequent calculations. 
We start by focusing on the equation of motion for the ground-state population $\rho_{00}$ [Eq.~\eqref{eq:rho_00}].
Setting $\dot{\rho}_{00}=0$, we find that the steady state satisfies
\begin{equation}
   \frac{\rho_{00}^{ss}}{\sum_{j=1}^N \rho_{jj}^{ss}} = \frac{k_d}{k_p}.
   \label{eq:ratio-00-exc}
\end{equation}
Combining this result with the conservation of probability, $\rho_{00}^{ss} + \sum_{j=1}^N \rho_{jj}^{ss}=1$, we obtain the total excitation probability and the ground-state population at steady state as
\begin{equation}
    \sum_{j=1}^N \rho_{jj}^{ss}=\frac{k_p}{k_p + k_d}
    \label{eq:tot-exc-prob}
\end{equation}
and 
\begin{equation}
    \rho_{00}^{ss}=\frac{k_d}{k_p + k_d},
    \label{eq:rho_00-ss}
\end{equation} 
respectively.
Both quantities are independent of the site energies ($E_j$) and intersite couplings ($V_{jl}$).
This means that varying the excited-state Hamiltonian ($H$) redistributes the steady-state population among the excited states while leaving the total population in the excited manifold, and therefore the ground-state population, unchanged.
\section{Steady state under strong coupling $V_{12}$}
\label{app:large-v12}
We evaluate the steady-state excitation probabilities in the limit $|V_{12}| \to \infty$ using density matrix perturbation theory~\cite{li_perturbative_2014}, which yields both the limiting values and the scaling of excitation probabilities with $|V_{12}|$. 
These asymptotic results are used in Appendix~\ref{app:rational} to determine the functional form of the input-output relation between steady-state excitation probabilities and $V_{12}$.

\subsection{Zeroth-order Liouvillian and perturbation}
\label{subapp:zero-perturb}
As $|V_{12}| \to \infty$, the couplings $V_{jl}$ between the strongly coupled sites $j \in A \equiv \{1,2\}$ and the remaining sites $l \in B \equiv \{3,\dots,N\}$ become negligible relative to $V_{12}$, effectively decoupling the two groups.
Since only site 1 is pumped [Eq.~\eqref{eq:rho_jj}], we expect only the $A$ sites to carry excitations at steady state.

To make this precise and capture the scaling of the $B$ excitation probabilities with $|V_{12}|$, we split the reduced dynamics of the subspace containing the ground-state populations, excited-state populations, and excited-state coherences as
\begin{equation}
    \dot{\tilde{\rho}} = (\mathscr{L}_0 + \mathscr{L}_1)\tilde{\rho},
\end{equation}
where $\tilde{\rho}$ is the corresponding reduced density matrix, the zeroth-order Liouvillian $\mathscr{L}_0\equiv \lim_{|V_{12}|\rightarrow\infty}\tilde{\mathcal{L}}$ retains only the $V_{12}$-dominated coherent dynamics of sites $A$ [Eqs.~\eqref{eq:rho_00}-\eqref{eq:rho_jl}],
\begin{align}
    (\mathscr{L}_0\tilde{\rho})_{00} &= k_{d}\sum_{j \in A\cup B}\rho_{jj}-k_{p}\rho_{00}, \nonumber\\
    (\mathscr{L}_0\tilde{\rho})_{11} &= k_{p}\rho_{00}-k_{d}\rho_{11}-i(V_{12}\rho_{21}-\rho_{12}V_{21}), \nonumber\\
    (\mathscr{L}_0\tilde{\rho})_{22} &= -k_{d}\rho_{22}-i(V_{21}\rho_{12}-\rho_{21}V_{12}), \nonumber\\
    (\mathscr{L}_0\tilde{\rho})_{12} &=-(k_{d}+k_{\phi})\rho_{12}-i(V_{12}\rho_{22}-\rho_{12}V_{21}),\nonumber\\
    (\mathscr{L}_0\tilde{\rho})_{21}&= (\mathscr{L}_0\tilde{\rho})_{12}^*,\nonumber\\
    (\mathscr{L}_0\tilde{\rho})_{jj} &= -k_{d}\rho_{jj}-i\sum_{\substack{m\in B \\ m\neq j}}(V_{jm}\rho_{mj}-\rho_{jm}V_{mj}),\quad j \in B, \nonumber\\
    (\mathscr{L}_0\tilde{\rho})_{jl} &=-\left[(k_{d}+k_{\phi})+i(E_{j}-E_{l})\right]\rho_{jl}\nonumber\\
    &\quad-i\left(\sum_{\substack{m\in B \\ m\neq j}}V_{jm}\rho_{ml}-\sum_{\substack{m\in B \\ m\neq l}}\rho_{jm}V_{ml}\right), \nonumber\\
    &\qquad\qquad\qquad\qquad\qquad j,l \in B,\ j \neq l,\nonumber\\
    (\mathscr{L}_0\tilde{\rho})_{1l} &=-(k_{d}+k_{\phi})\rho_{1l}-iV_{12}\rho_{2l},\quad l \in B,\nonumber\\
     (\mathscr{L}_0\tilde{\rho})_{2l} &=-(k_{d}+k_{\phi})\rho_{2l}-iV_{21}\rho_{1l},\quad l \in B,\nonumber\\
     (\mathscr{L}_0\tilde{\rho})_{jl} &=(\mathscr{L}_0\tilde{\rho})_{lj}^*,\quad j\in B,\ l\in A,
     \label{eq:L0}
\end{align}
and the perturbation $\mathscr{L}_1$ restores the $A$-$B$ couplings, 
\begin{align}
    &(\mathscr{L}_1\tilde{\rho})_{jl} =\nonumber\\
    &\begin{cases}
        0, & j,l=0 \\
        -i\sum_{m\in B}(V_{jm}\rho_{ml}-\rho_{jm}V_{ml}), & j,l\in A \\
        -i\sum_{m\in A}(V_{jm}\rho_{ml}-\rho_{jm}V_{ml}), & j, l \in B\\
        -i\left(\sum_{m\in B}V_{jm}\rho_{ml}-\sum_{m\in A}\rho_{jm}V_{ml}\right), & j\in A,\ l\in B\\
        -i\left(\sum_{m\in A}V_{jm}\rho_{ml}-\sum_{m\in B}\rho_{jm}V_{ml}\right), & j\in B,\ l\in A
    \end{cases}
    \label{eq:L1}
\end{align}
which indirectly couples $A$-exclusive and $B$-exclusive density matrix elements through the $A$-$B$ coherences.
\subsection{Zeroth-order steady state}
The zeroth-order steady state, which satisfies
\begin{equation}
    \mathscr{L}_0 \tilde{\rho}^{ss(0)} =0,
\end{equation}
has the solution 
\begin{equation}
    \tilde{\rho}_{jl}^{ss(0)}=
    \begin{cases}
        \frac{k_d}{k_p + k_d}, & j,l=0 \\
        \frac{k_p}{2(k_p + k_d)}, & j=l \in A \\
        0, & \text{else}
    \end{cases}.
    \label{eq:rho-ss-0}
\end{equation}
This confirms that, as $|V_{12}|\rightarrow\infty$, all excitation is confined to sites $A$, shared equally between sites 1 and 2, with no surviving coherences. 
\subsection{Properties of the pseudoinverse of the zeroth-order Liouvillian}
\label{subapp:pseudoinverse}
We collect properties of $\mathscr{L}_0^+$, the Moore-Penrose pseudoinverse of the zeroth-order Liouvillian $\mathscr{L}_0$, needed to evaluate the perturbative corrections.
We partition the density matrix into blocks,
\begin{equation}
    \tilde{\rho} =
    \begin{pmatrix}
        \rho_{GA} & \rho_{X}\\
        \rho_{X}^{\dagger} & \rho_{B}
    \end{pmatrix},
    \label{eq:rho-tilde-ab}
\end{equation}
where $\rho_{GA}\equiv \{\rho_{00}\} \cup \{\rho_{jl}\}_{j,l \in A}$ contains the ground-state population and $A$-site density matrix elements; $\rho_{B}\equiv \{\rho_{jl}\}_{j,l \in B}$ contains the $B$-site density matrix elements; and $\rho_{X}\equiv \{\rho_{jl}\}_{j \in A,\, l \in B}$ contains the $A$-$B$ cross coherences.
The zeroth-order Liouvillian inherits the block structure
\begin{equation}
    \mathscr{L}_0 =
    \begin{pmatrix}
        \mathcal{L}_{GAB} & \mathbf{0}\\
        \mathbf{0} & \mathcal{L}_{X}
    \end{pmatrix},
    \label{eq:L0-block}
\end{equation}
where
\begin{equation}
    \mathcal{L}_{GAB} \equiv
    \begin{pmatrix}
        \mathcal{L}_{GA,GA} & \mathcal{L}_{GA,B} \\
        \mathbf{0} & \mathcal{L}_{B,B}
    \end{pmatrix},
    \label{eq:L-GAB}
\end{equation}
governs the ground state, $A$, and $B$ density matrix elements, while $\mathcal{L}_{X}$ governs the $A$-$B$ cross coherences (fully decoupled from the other density matrix elements at zeroth order).
Here, $\mathbf{0}$ denotes a matrix whose entries are all 0 and dimensionality can be inferred from that of the neighboring nonzero blocks.
The pseudoinverse is correspondingly block diagonal,
\begin{equation}
    \mathscr{L}_0^+ =
    \begin{pmatrix}
        \mathcal{L}_{GAB}^+ & \mathbf{0}\\
        \mathbf{0} & \mathcal{L}_{X}^+
    \end{pmatrix},
    \label{eq:L0-plus}
\end{equation}

\paragraph{Block $\mathcal{L}_X^+$.}
Since the cross coherences are purely decaying (no coupling to populations in $\mathscr{L}_0$), $\mathcal{L}_X$ is invertible:
\begin{equation}
    \mathcal{L}_X^+ = \mathcal{L}_X^{-1} =\bigoplus_{l \in B}(\mathcal{L}_{Al}^{-1} \oplus \mathcal{L}_{lA}^{-1})
    \label{eq:L-X-plus}
\end{equation}
where 
\begin{equation}
    \mathcal{L}_{Al}^{-1} =
    \frac{1}{V_{12}^2 + (k_d + k_\phi)^2}
    \begin{pmatrix}
        -(k_d + k_\phi) & iV_{12} \\
        iV_{21} & -(k_d + k_\phi)
    \end{pmatrix}
\end{equation}
acts on the coherences $\rho_{jl}$ for $j\in A$ and $l\in B$, while $\mathcal{L}_{lA}^{-1} = (\mathcal{L}_{Al}^{-1})^*$ acts on $\rho_{lj} = \rho_{jl}^*$.

\paragraph{Block $\mathcal{L}_{GAB}^+$.}
Using the pseudoinverse formulas of~\cite{miao_general_1991} for $2\times 2$ block matrices, we establish two key properties.
First, the rows of $\mathcal{L}_{GAB}^+$ corresponding to $B$ sites are independent of $V_{12}$.
Second, the part of these rows representing the coupling of the $A$ populations and coherences to the corresponding $B$ elements satisfies
\begin{align}
    (\mathcal{L}_{GAB}^+)_{jl,11} &= (\mathcal{L}_{GAB}^+)_{jl,22}, \nonumber\\
    (\mathcal{L}_{GAB}^+)_{jl,12} &= (\mathcal{L}_{GAB}^+)_{jl,21}=0,
    \label{eq:props-L-GAB}
\end{align}
for $j,l \in B$. 
Both properties follow from the fact that $\mathcal{L}_{GA,GA}$, explicitly [Eq.~\eqref{eq:L0}]
\begin{align}
    &\mathcal{L}_{GA,GA} = \nonumber\\
    &
    \begin{pmatrix}
        -k_{p} & k_{d} & k_{d} & 0 & 0\\
        k_{p} & -k_{d} & 0 & iV_{21} & -iV_{12}\\
        0 & 0 & -k_{d} & -iV_{21} & iV_{12}\\
        0 & iV_{12} & -iV_{12} & -(k_{d}+k_{\phi}) & 0\\
        0 & -iV_{21} & iV_{21} & 0 & -(k_{d}+k_{\phi}),
        \end{pmatrix}
\end{align}
(rows/columns ordered as $\rho_{00},\rho_{11},\rho_{22},\rho_{12},\rho_{21}$), admits a unique steady state.
This can be seen from the projector onto the null space of $\mathcal{L}_{GA,GA}^\dagger$,
\begin{equation}
    \mathbf{I}-\mathcal{L}_{GA,GA}\mathcal{L}_{GA,GA}^{+}= 
    \begin{pmatrix}
        \frac{1}{3} & \frac{1}{3} & \frac{1}{3} & 0 & 0\\
        \frac{1}{3} & \frac{1}{3} & \frac{1}{3} & 0 & 0\\
        \frac{1}{3} & \frac{1}{3} & \frac{1}{3} & 0 & 0\\
        0 & 0 & 0 & 0 & 0\\
        0 & 0 & 0 & 0 & 0
    \end{pmatrix},
    \label{eq:proj-L-GA}
\end{equation}
which projects onto the identity density matrix (equal populations, zero coherences), confirming a one-dimensional null space.
Consequently, the projection $\mathbf{P}\equiv(\mathbf{I}-\mathcal{L}_{GA,GA}\mathcal{L}_{GA,GA}^{+})\mathcal{L}_{GA,B}$ has rows that are either zero or a multiple of the single nonzero row of $\mathcal{L}_{GA,B}$, so $\mathcal{L}_{GA,B}(\mathbf{I} -\mathbf{P}^+ \mathbf{P})=\mathbf{0}$.
Since $\mathbf{P}$ is independent of $V_{12}$, the formulas of~\cite{miao_general_1991} imply that the $B$-site rows of $\mathcal{L}_{GAB}^+$ are $V_{12}$-independent, establishing the first property.
For the second property, since $\mathcal{L}_{GA,GA}$ is a Lindblad generator with a unique steady state, the null space of $\mathcal{L}_{GA,GA}^\dagger$ is spanned by the identity density matrix, which has support only on populations, thus implying Eq.~\eqref{eq:props-L-GAB}.
\subsection{Perturbative corrections to the steady state}
The $k$th-order correction to the steady state satisfies the recursion~\cite{li_perturbative_2014}
\begin{equation}
     \tilde{\rho}^{ss(k)} = -\mathscr{L}_0^+ \mathscr{L}_1 \tilde{\rho}^{ss(k-1)}.
\end{equation}

At first order, $\mathscr{L}_1$ converts $A$-site excitation probabilities in $\tilde{\rho}^{ss(0)}$ into $A$-$B$ coherences,
\begin{equation}
    (\mathscr{L}_1\tilde{\rho}^{ss(0)})_{jl}=\frac{k_p}{2(k_d + k_p)} \begin{cases}
        i V_{jl}, & j\in A,\ l\in B \\
        -i V_{jl}, & j\in B,\ l\in A \\
        0, & \text{else}
    \end{cases}.
\end{equation}
Applying $-\mathscr{L}_0^+$ mixes coherences $\rho_{1l}$ and $\rho_{2l}$ (and their conjugates) via $V_{12}$, giving 
\begin{align}
    \tilde{\rho}^{ss(1)}_{jl} &=(-\mathscr{L}_0^+ \mathscr{L}_1\tilde{\rho}^{ss(0)})_{jl} \\
    &=\frac{k_p}{2(k_d + k_p)[V_{12}^2 + (k_d + k_\phi)^2]} \nonumber\\
    &
    \quad \times
    \begin{cases}
        i(k_d + k_\phi)V_{1l} + V_{12}V_{2l}, & j=1,\ l\in B \\
        i(k_d + k_\phi)V_{2l}+V_{21}V_{1l}, & j=2,\ l\in B \\
        -iV_{j1}(k_d + k_\phi) + V_{j2}V_{21}, & j\in B,\ l=1 \\
        -iV_{j2}(k_d + k_\phi)+V_{j1}V_{12}, & j\in B,\ l=2 \\
        0, & \text{else}
    \end{cases}.
\end{align}
No $B$-site excitation probability appears at this order. 

At second order, applying $\mathscr{L}_1$ to $\tilde{\rho}^{ss(1)}$ generates $B$-site populations and coherences, as well as corrections to the corresponding density matrix elements of the $A$ sites:
\begin{align}
    &(\mathscr{L}_1\tilde{\rho}^{ss(1)})_{jl} \nonumber\\
    &= -\frac{k_p(k_d + k_\phi)}{(k_d + k_p)[V_{12}^2 + (k_d + k_\phi)^2]}\nonumber\\
    &\quad\times
    \begin{cases}
        \sum_{m\in B} V_{1m}^2, & j=l=1 \\
        \sum_{m\in B} V_{2m}^2, & j=l=2 \\
        \sum_{m\in B}\left(V_{1m}V_{m2} + \frac{iV_{12}(V_{1m}^2 - V_{2m}^2)}{2(k_d + k_\phi)}\right), & 
        \begin{matrix}
            j=1, \\
            l=2 
        \end{matrix} \\
        \sum_{m\in B}\left(V_{2m}V_{m1} + \frac{iV_{21}(V_{2m}^2 - V_{1m}^2)}{2(k_d + k_\phi)}\right), & 
        \begin{matrix}
            j=2, \\
            l=1 
        \end{matrix} \\
        \sum_{m\in A} V_{jm} V_{ml}, & j,l \in B \\
        0, & \text{else}
    \end{cases}
    \label{eq:L1-rho-ss-1}
\end{align}
where, in evaluating the expressions for $j=l=1,2$ (first two cases), we have used the fact that all $V_{jl}$ are real-valued.
The second-order correction to the $B$-site excitation probabilities is then
\begin{align}
    \left. \tilde{\rho}_{jj}^{ss(2)}\right|_{j\in B}&= -(\mathscr{L}_0^+ \mathscr{L}_1 \rho^{ss(1)})_{jj}\\
    &= -\sum_{l,m \in A} (\mathscr{L}_0^+)_{jj,lm} (\mathscr{L}_1 \tilde{\rho}^{ss(1)})_{lm}\nonumber\\
    &\quad
    -\sum_{l,m \in B} (\mathscr{L}_0^+)_{jj,lm} (\mathscr{L}_1 \tilde{\rho}^{ss(1)})_{lm},
\end{align}
Using Eq.~\eqref{eq:props-L-GAB}, which states that the $B$-site rows of $(\mathscr{L}_0^+)$ treat sites 1 and 2 symmetrically and do not couple to $A$ coherences, together with Eq.~\eqref{eq:L1-rho-ss-1}, we find 
\begin{equation}
    \tilde{\rho}_{jj}^{ss(2)}\sim \frac{1}{V_{12}^2},\quad j\in B,
\end{equation}
where we have applied the limit $V_{12}\rightarrow\infty$.
Since this is the lowest-order correction to $\tilde{\rho}_{jj}^{ss}$ for $j\in B$, we conclude that the $B$ excitation probabilities decay as $V_{12}^{-2}$ as $|V_{12}|\rightarrow\infty$.

\section{Rational form of input-output-relation}
\label{app:rational}
Here, we derive the rational form~\eqref{eq:rational-main}-\eqref{eq:degree-main} of the input-output relation $p_i(V_{12})\equiv \rho_{ii}^{ss}(V_{12})$ for $i=1,\dots,N$, or the excited-state populations at site $i$ as a function of the coupling between sites 1 and 2.

We begin by flattening the density matrix $\rho$ to a vector and, correspondingly, the Liouvillian $\mathcal{L}$ to a matrix.
Specifically, we treat the index pair $jl$ of the density matrix element $\rho_{jl}$ as a single flattened index, which makes $\mathcal{L}_{jl,mn}$ the element in row $jl$ and column $mn$ of the Liouvillian matrix. 

We next focus on the Liouvillian $\tilde{\mathcal{L}}$ that describes the dynamics of the subspace in which the excited-state populations evolve together with the ground-state population and the excited-state coherences (Appendix~\ref{app:weak-pumping}).
Since the steady state of this subspace lies in the nullspace of $\tilde{\mathcal{L}}$, it is proportional to a column $jl$ of the adjugate matrix~$\mathrm{adj}(\tilde{\mathcal{L}})$~(\cite{horn_matrix_2012}, Eq. 0.8.2.1).
The adjugate element $[\mathrm{adj}(\tilde{\mathcal{L}})]_{jl,mn}$ denotes the cofactor associated with row $mn$ and column $jl$ of $\tilde{\mathcal{L}}$.
The steady state is unique because decay provides a relaxation pathway from every excited state to the ground state, preventing dynamically disconnected subspaces that could support multiple steady states. Therefore, the columns of $\mathrm{adj}(\tilde{\mathcal{L}})$ differ from each other only by a global prefactor, implying that the steady state is, in fact, proportional to \textit{any} column $jl$ of $\mathrm{adj}(\tilde{\mathcal{L}})$.
The conservation of probability, $\text{Tr}[\rho]=1$, sets the normalization of the steady state.
With these facts in mind, the steady-state population serving as the computational output can be written as
\begin{equation}
    p_i(V_{12})
    =
    \frac{
    [\mathrm{adj}(\tilde{\mathcal{L}})]_{ii,jl}
    }{
    \sum_{m=0}^{N}
    [\mathrm{adj}(\tilde{\mathcal{L}})]_{mm,jl}
    }
    \label{eq:rho-adj}
\end{equation}
for any choice of the flattened index $jl$. 

We are now ready to show that $p_i$ is a rational function of $V_{12}$. 
We notice that $V_{12}$ enters the Liouvillian $\tilde{\mathcal{L}}$ only through the matrix elements
$\tilde{\mathcal{L}}_{1j,2j}=-iV_{12}$,
$\tilde{\mathcal{L}}_{2j,1j}=-iV_{21}$,
$\tilde{\mathcal{L}}_{j1,j2}=iV_{21}$, and
$\tilde{\mathcal{L}}_{j2,j1}=iV_{12}$ for $j=1,\ldots,N$.
Since each element of $\mathrm{adj}(\tilde{\mathcal{L}})$ is the determinant of a submatrix of $\tilde{\mathcal{L}}$, every numerator and denominator appearing in Eq.~\eqref{eq:rho-adj} is polynomial in $V_{12}$.
Therefore, by fixing the same $jl$ in Eq.~\eqref{eq:rho-adj} for all $i$, we find that $p_i(V_{12})$ is a rational function whose form is given by Eq.~\eqref{eq:rational-main}.

We proceed to determine the degree of the numerator polynomials $P_i$ and denominator polynomial $Q$ [Eq.~\eqref{eq:rational-main}].
To do so, we hereafter make the association
\begin{equation}
    P_i(V_{12}) \equiv [\mathrm{adj}(\tilde{\mathcal{L}})]_{ii,00},
\end{equation}
corresponding to the choice $jl=00$ in Eq.~\ref{eq:rho-adj}.
Using this convention in Eq.~\eqref{eq:rho-adj}, it immediately follows that
\begin{equation}
    Q(V_{12}) \equiv  [\mathrm{adj}(\tilde{\mathcal{L}})]_{00,00} + \sum_{i=1}^N P_i (V_{12}).
    \label{eq:q0}
\end{equation}

Focusing first on $P_1$, which corresponds to the pumped site ($i=1$), it will be convenient to adopt the alternative expression
\begin{equation}
    P_{1}(V_{12}) = \zeta\det(\tilde{\mathcal{L}}[I,J]),
\end{equation}
where we have recast the submatrix $\tilde{\mathcal{L}}^{[00,11]}$ of $\tilde{\mathcal{L}}$ as $\tilde{\mathcal{L}}[I,J]$, which denotes keeping only rows $I=\{jl\}_{j,l=1}^N$ and columns $J=\{00\}\cup\{jl\}_{j,l=1}^N\backslash\{11\}$.  
We have also defined the sign associated with $[\text{adj}(\tilde{\mathcal{L}})]_{11,00}$ as $\zeta \equiv (-1)^{\eta(00)+\eta(11)}$, where $\eta:\{jl\}\rightarrow \{1,\dots,N^2 + 1\}$ is a 1-to-1 map that indicates the order of each index $jl$ of the density matrix after flattening.

Next, we use the Leibniz formula to expand the determinant as 
\begin{align}
P_{1}(V_{12})=\zeta\sum_{f}\text{sgn}(f)\prod_{jl\in I}(\tilde{\mathcal{L}}[I,J])_{jl,f(jl)},
\label{eq:leibniz}
\end{align}
where the sum runs over all one-to-one maps $f:I\rightarrow J$ of indices, and $\text{sgn}(f)\in\{\pm1\}$ is determined by the Leibniz formula. 
By definition of $f$, we can view each term in the sum as a network of transitions from a row index $jl\in I$ to a column index $mn \in J$, where each $jl$ serves as the starting state for a unique transition, each $mn$ serves as the final state for a unique transition, and the transition $jl\rightarrow mn$ is associated with the matrix element $(\tilde{\mathcal{L}}[I,J])_{jl,mn}$. 
The value of each term in the sum, up to a sign, i.e., $\text{sgn}(f)$, is then given by the product of the matrix elements corresponding to all transitions. 

Thus, determining the degree of $P_1$ is equivalent to identifying the maximum number of times $V_{12}$ (or $V_{21}=V_{12}^*$) can appear in the aforementioned networks such that the terms containing the maximum number do not completely cancel each other out. 
By solving the restated problem, we find that $P_1$ is a polynomial of degree $4N-6$ whose leading term is
\begin{widetext}
\begin{align}
    -\zeta \det(\tilde{\mathcal{L}}[I',I'])\sum_{jl\in\{12,21\}}\tilde{\mathcal{L}}_{11,00}\tilde{\mathcal{L}}_{jl,22}\tilde{\mathcal{L}}_{22,jl}\tilde{\mathcal{L}}_{lj,lj}\left(\prod_{j=3}^{N}\tilde{\mathcal{L}}_{j1,j2}\tilde{\mathcal{L}}_{j2,j1}\tilde{\mathcal{L}}_{2j,1j}\tilde{\mathcal{L}}_{1j,2j}\right)=-2\zeta\det(\tilde{\mathcal{L}}[I',I'])k_{p}(k_{d}+k_{\phi})V_{12}^{4N-6},
    \label{eq:leading-p1-v12}
\end{align}
\end{widetext}
where $I'=\{jl\}_{j,l=3}^N$ is the set of indices of the populations and coherences involving only excited states $j,l=3,\dots,N$. 
We note that $\det(\tilde{\mathcal{L}}[I',I'])\neq 0$, or equivalently, $\tilde{\mathcal{L}}[I',I']$ has a trivial null space, since $\tilde{\mathcal{L}}[I',I']$ acts on density matrix elements that do not include the ground state and thus decay to 0 over time.
A graph representation of the $jl=12$ term of Eq.~\eqref{eq:leading-p1-v12} is shown in Fig.~\ref{fig:quantum-expressivity}d.

The degrees of $P_i$ for $i=2,\dots,N$ follow from the large-$|V_{12}|$ limit: as $V_{12}\rightarrow\pm\infty$, sites 1 and 2 effectively decouple from the rest of the network.
In this regime, $p_1$ and $p_2$ remain finite, while $p_i\rightarrow0$ for $i\geq 3$, with $p_i\sim V_{12}^{-2}$ (Appendix~\ref{app:large-v12}), yielding the stated scaling of $\deg P_i$ in Eq.~\eqref{eq:degree-main}. 

We finally turn to the degree of the denominator polynomial $Q$ [Eq.~\eqref{eq:rational-main}].
We recall Eq.~\eqref{eq:q0}, which expresses $Q$ in terms of the numerator polynomials $P_i$ and $[\mathrm{adj}(\tilde{\mathcal{L}})]_{00,00}$.   
Using the normalization of total excitation probability, Eq.~\eqref{eq:tot-exc-prob}, which holds for all $V_{12}$, we can eliminate $[\mathrm{adj}(\tilde{\mathcal{L}})]_{00,00}$ from Eq.~\eqref{eq:q0}, yielding 
\begin{equation}
Q(V_{12})
=
\frac{k_p+k_d}{k_p}
\sum_{i=1}^{N}
P_i(V_{12}).
\label{eq:q}
\end{equation}
Thus, by the degrees of the $P_i$ polynomials, we obtain $\deg Q = 4N-6$, as given in Eq.~\eqref{eq:degree-main}.
\section{Incoherent regime}
\label{app:incoherent}
Here, we consider the incoherent regime, in which the dephasing rate goes to infinity, $k_\phi \rightarrow \infty$.
In this regime, all coherences are fully suppressed, so that the populations become dynamically decoupled from the coherences. 
In Sec.~\ref{subapp:dynamics-incoherent}, we follow~\cite{cao_optimization_2009,hoyer_limits_2010} to show that the population dynamics in the incoherent regime reduces to the kinetic model~\eqref{eq:kinetics}.
In Sec.~\ref{subapp:rational-incoherent}, we derive the rational form of the input-output relation $p_i(V_{12})$ ($i=1,\dots,N$) as given by Eqs.~\eqref{eq:rational-main} and~\eqref{eq:p-incoherent}. 
\subsection{Population dynamics reduces to kinetic model}
\label{subapp:dynamics-incoherent}
Here, we show that the population dynamics in the incoherent regime reduces to kinetic model~\eqref{eq:kinetics}.
Specifically, we extend the derivation of~\cite{cao_optimization_2009,hoyer_limits_2010} for short-range coupling to the all-to-all coupling considered here.
This result implies that the populations become dynamically decoupled from all coherences. 

From Appendix~\ref{app:weak-pumping}, in particular Eqs.~\eqref{eq:rho_00}-\eqref{eq:rho_jl}, we recall that the populations are dynamically coupled to only the excited-state coherences, $\rho_{jl}$ for $j,l=1,\dots, N$.
In the incoherent regime, $k_\phi \rightarrow \infty$, the coherences decay so fast that they can be approximated as a constant on the time scale of the evolution of the populations (known as steady-state approximation~\cite{atkins_physical_2006} or adiabatic elimination~\cite{brion_adiabatic_2007}),
\begin{equation}
\dot{\rho}_{jl}\approx0, \quad j,l=1,\dots, N,
\end{equation}
which yields
\begin{align}
    \rho_{jl}\approx&-\frac{i}{(k_{d}+k_{\phi})+i(E_{j}-E_{l})}\nonumber\\
    &\quad\times\left(\sum_{m\neq j}V_{jm}\rho_{ml}-\sum_{m\neq l}\rho_{jm}V_{ml}\right).
\end{align}
We see that the coherences contribute a factor of $O(1/k_{\phi})$.
Since we are working in the limit of large $k_{\phi}$, we can further drop the coherences on the right-hand side:
\begin{equation}
\rho_{jl}\approx-\frac{iV_{jl}}{(k_{d}+k_{\phi})+i(E_{j}-E_{l})}(\rho_{ll}-\rho_{jj}).
\end{equation}
Substituting this result into Eq.~\eqref{eq:rho_jj}, we find that the equations of motion for the populations, Eqs.~\eqref{eq:rho_00}-\eqref{eq:rho_jj}, reduce to the kinetic model~\eqref{eq:kinetics}.
\subsection{Rational form of input-output relation}
\label{subapp:rational-incoherent}
Here, we show that the input-output relation $p_i (V_{12}) \equiv \rho_{ii}^{ss}(V_{12})$ in the incoherent regime $k_\phi \rightarrow \infty$ is a rational function of the form given by Eqs.~\eqref{eq:rational-main} and~\eqref{eq:p-incoherent}.
We closely follow the derivation (see main text) of the corresponding result for general parameters [Eqs.~\eqref{eq:rational-main}-\eqref{eq:degree-main}]. 

First, we rewrite the effective kinetic model~\eqref{eq:kinetics} describing the population dynamics in the incoherent regime as
\begin{equation}
\dot{\rho}_{jj}=\sum_{l=0}^{N}W_{jl}\rho_{ll},
\end{equation}
where 
\begin{equation}
    W_{jl}=
    \begin{cases}
        k_{lj}, & j\neq l,\\
        -\sum_{m\neq j}k_{jm}, & j=l,
    \end{cases}
\label{eq:w}
\end{equation}
is the transition-rate matrix. 
Using the steady-state relation $\sum_l W_{jl}\rho_{ll}^{ss} = 0$ for all $j$, we can express the computational output $p_i$ in terms of the adjugate matrix $\mathrm{adj}(W)$ of $W$ as~(\cite{horn_matrix_2012}, Eq. 0.8.2.1)
\begin{equation}
    p_i(V_{12})=\frac{[\mathrm{adj}(W)]_{i,j}}{\sum_{l=0}^{N}[\mathrm{adj}(W)]_{l,j}},
    \label{eq:rho-adj-incoherent}
\end{equation}
where $j$ can take any value due to the uniqueness of the steady state.
The form of the denominator reflects the conservation of total probability, $\sum_{i=0}^N p_i=1$. 
From Eqs.~\eqref{eq:w} and~\eqref{eq:k_jl}, we notice that the only elements of $W$ which depend on $V_{12}$ are $W_{12}$, $W_{21}$, $W_{11}$, and $W_{22}$; in particular, they go as $\sim V_{12}^{2}$.
Since each element of $\mathrm{adj}(W)$ is the determinant of a submatrix of $W$, then every numerator and denominator in Eq.~\eqref{eq:rho-adj-incoherent} is a polynomial in $V_{12}$.
Thus, by fixing the same $j$ in Eq.~\eqref{eq:rho-adj-incoherent} for all $i$, we see that the outputs $p_i$ are rational functions of the input $V_{12}$ whose form is given by Eq.~\eqref{eq:rational-main}.

To determine the degrees of the numerator polynomials $P_i$ and denominator polynomial $Q$, we first make the association  
\begin{equation}
    P_i(V_{12}) \equiv[\mathrm{adj}(W)]_{i,0},
\end{equation}
corresponding to the choice $j = 0$ in Eq.~\eqref{eq:rho-adj-incoherent}. 
We now focus on the polynomial for the pumped site, which can be expressed as
\begin{equation}
    P_{1} = -\det(W[I, J]),
\end{equation}
where $W[I, J]$ is the submatrix of $W$ obtained by keeping only rows $I = \{j\}_{j=1}^N$ and columns $J=\{0\}\cup\{j\}_{j=2}^N$.
Following Appendix~\ref{app:rational}, we employ the Leibniz expansion of the determinant [see Eq.~\eqref{eq:leibniz}] and find that $P_1$ takes the form of a polynomial with degree 2, 
\begin{equation}
    P_1(V_{12})=m_1 V_{12}^2 + b_1,
    \label{eq:p-1-incoherent}
\end{equation}
where
\begin{equation}
    m_1=\frac{2\det(W[I',I'])k_p (k_d + k_\phi)}{(k_{d}+k_{\phi})^{2}+(E_{1}-E_{2})^{2}},
\end{equation}
$I'=\{3,\dots,N\}$ and $b_1$ is a constant. 
We note that $\det(W[I',I'])\neq 0$, or equivalently, $W[I',I']$ has a trivial null space, since $W[I',I']$ acts on states that do not include the ground state and thus decay to 0 over time.
Using $\deg P_1 = 2$, the asymptotic behavior of $p_i(V_{12})$ as $|V_{12}|\rightarrow\infty$ (Appendices~\ref{app:large-v12} and~\ref{app:rational}), and the rational form~\eqref{eq:rational-main} of $p_i$, we find that the numerator polynomials for the other sites take the form
\begin{align}
    P_2(V_{12})&=m_2 V_{12}^2 + b_2,\\
    P_i(V_{12})&=b_i,\quad i=3,\dots,N,
    \label{eq:p-not1-incoherent}
\end{align}
where $m_2$ and $b_2,\dots,b_N$ are constants. 
Knowing that the probabilities must be nonnegative for all $V_{12}$, it can be shown that $m_2$ and all $b_i$ have the same sign as $m_1$.
This result generally holds, except when any of the sites are not coupled at all to site 1, either directly or indirectly through another site, in which case the $b_i$ corresponding to each decoupled site vanishes.
With these sign relations, Eqs.~\eqref{eq:p-1-incoherent} and~\eqref{eq:p-not1-incoherent} correspond to Eq.~\eqref{eq:p-incoherent} of the main text.
Accounting for the value of the total excitation probability, as discussed in the main text (Sec.~\ref{sec:quantum-expressivity}) for the case of general parameters, we can write the denominator polynomial $Q$ in the form of Eq.~\eqref{eq:q}.
\section{Numerical details}
\label{app:numerical}
\subsection{Random networks}
\label{subapp:random-networks}
We generate random excitonic networks with $N$ sites by sampling the excitation energies $E_j$ and couplings $V_{jl}$ as independently and identically distributed (i.i.d.) random variables from uniform distributions, $E_j\sim U(\bar{E} - \delta E, \bar{E} + \delta E)$ for $\delta E = 0.5/\sqrt{N}$ and $V_{jl}\sim U(-V_{\max}, V_{\max})$ for $V_{\max} = 1/\sqrt{N}$.
We choose $\bar{E}=100$ so that the excitation energies are much larger than the energy $E_0 = 0$ chosen for the vacuum state. 
The pumping and decay rates are $k_p = 0.001/\sqrt{N}$ and $k_d = 0.01/\sqrt{N}$, respectively.
The dephasing rate is $k_\phi=0.01/\sqrt{N}$ in Fig.~\ref{fig:quantum-expressivity}.
For the calculations using the Redfield form of dephasing (Fig.~\ref{fig:temperature-dependence}), we use an Ohmic spectral density [Eq.~\eqref{eq:j-ohmic}] with prefactor $\gamma=0.01$ and cutoff frequency $\Omega_\text{cut}=2.0$.
All energy differences, couplings, and dissipation rates are scaled by $1/\sqrt{N}$ so that the expected energy bandwidth $\Delta E = \max_\alpha E_\alpha - \min_\alpha E_\alpha$ of the exicted eigenstates ($\alpha$) and the relative time scale of the coherent and dissipative dynamics do not change with $N$. 
Indeed, as $N$ increases, Wigner's semicircle law gives $\Delta E \rightarrow (4/\sqrt{3})V_{\max}\sqrt{N}$, which is finite since $V_{\max}\propto 1/\sqrt{N}$. 
Throughout the main text, the results represent an ensemble of 4900 random networks per parameter set. 
\subsection{Steady-state computation}
\label{subapp:steady-state}
We numerically evaluate the input-output relation $p_i(V_{12})\equiv \rho^{ss}_{ii}$ using QuTiP~\cite{lambert_qutip_2026} to construct the Liouvillian [$\mathcal{L}$, first line of Eq.~\eqref{eq:lindblad}] and directly solve for the steady state.
In calculations involving the Redfield dephasing term [Eq.~\eqref{eq:redfield-dephasing}], we do not apply the secular approximation~\cite{breuer_theory_2002,may_charge_2011} to that term.
Analysis of the input-output relation involves 511 values of the output computed at input values in the following intervals with their corresponding grid sizes, which reflects the energies and couplings of the random excitonic networks [Appendix~\ref{subapp:random-networks}]:
\begin{align}
    &V_{12}\in [-4, -3),\; \Delta V_{12}= 0.2,\nonumber\\
    &V_{12}\in [-3, -2),\; \Delta V_{12}= 0.1,\nonumber\\
    &V_{12}\in [-2, -1),\; \Delta V_{12}= 0.025,\nonumber\\
    &V_{12}\in [-1, 1),\; \Delta V_{12}= 0.005,\nonumber\\
    &V_{12}\in [1, 2),\; \Delta V_{12}= 0.025,\nonumber\\
    &V_{12}\in [2, 3),\; \Delta V_{12}= 0.1,\nonumber\\
    &V_{12}\in [3, 4],\; \Delta V_{12}= 0.2.
    \label{eq:input-values}
\end{align}
\subsection{Fitting input-output relations to rational functions}
\label{subapp:fitting}
For the Lindblad steady state, we numerically verify that the input-output relation $p_i(V_{12})$ for  $i=1,\dots,N$ is a rational function of the form \eqref{eq:rational-main}-\eqref{eq:degree-main} by fitting each $p_i$ to the rational ansatz 
\begin{align}
    p_i(V_{12}) = \frac{\sum_{s=0}^d a_{is} T_s(V_{12})} {[(k_p + k_d)/k_p]\sum_{j=1}^N \sum_{s=0}^d a_{js}T_s(V_{12})},
    \label{eq:ansatz}
\end{align}
where $\{a_{is}\}$ are the fitting coefficients, the prefactor $(k_p + k_d) / k_p$ in the denominator reflects the analytically derived denominator~\eqref{eq:q}, and both the numerator and denominator are polynomials of degree $d$ expressed in terms of the Chebyshev polynomials of the first kind, $T_s$.
The fitting is carried out using values of the output $\{p_i(V_{12})\}$ evaluated at $N_\text{in}=401$ uniformly spaced input values $V_{12}\in [-1, 1]$ (i.e., spacing $\Delta V_{12} = 0.005$).  
To facilitate the fitting, we enforce that the constant terms add up to 1, $\sum_{j=1}^N a_{j0} = 1$, though any nonzero value should have the same effect because Eq.~\eqref{eq:ansatz} is invariant to global rescaling of $\{a_{is}\}$.
To determine the best fit, we recast the input-output relations at all $N$ sites and all $N_\text{in}$ input values, along with the constraint on the constant terms, as a system of $N_\text{in}N+1$ equations that are linear in $\{a_{is}\}$.
We then solve for $\{a_{is}\}$ using least squares, specifically, the GELSY driver in the \texttt{scipy.linalg.lstsq} routine of SciPy \cite{scipy_2020}. 
Fig.~\ref{fig:quantum-expressivity} shows the relative error between the predicted and actual outputs at the input values listed in Eq.~\eqref{eq:input-values}.
\section{Parity of turning points}
\label{app:parity}
The turning-point counts in the input-output relations of random excitonic networks (Figs.~\ref{fig:quantum-expressivity}f,~\ref{fig:coherent-advantage}d, and~\ref{fig:temperature-dependence}b) are predominantly odd.
This can be understood from the asymptotic behavior of $p_i(V_{12})$ as $V_{12}\rightarrow\pm\infty$ (Sec.~\ref{app:large-v12}).
In this limit, the eigenstates of $H$ [Eq.~\eqref{eq:h}], their energies, and their site probabilities are virtually identical for $V_{12}\rightarrow+\infty$ and $V_{12}\rightarrow-\infty$, differing only in the sign of the amplitude on sites 1 and 2 for the highest and lowest eigenstates.
Since excitation probabilities are insensitive to this sign change, $p_i(V_{12})$ approaches the same asymptotic value from the same direction as $V_{12}\rightarrow+\infty$ and $V_{12}\rightarrow-\infty$, as follows from the leading asymptotic correction being even in $V_{12}$ (Appendix~\ref{app:large-v12}).
Because $p_i(V_{12})$ is a rational function of finite degree (Sec.~\ref{sec:quantum-expressivity}), it is eventually monotone in both tails.
A function that is monotone in both tails and approaches the same value from the same direction necessarily has an odd number of turning points.
% TODO: create figure
For sites $i=3,\dots,N$, odd parity in fact holds exactly: since these populations approach zero as $|V_{12}|\rightarrow\infty$ (Appendix~\ref{app:large-v12}) and are non-negative, they must approach from above on both sides, regardless of subleading corrections.
Even counts can therefore arise only when the output site is 1 or 2, whose asymptotic populations are finite and thus not constrained by non-negativity.
The rare even counts likely arise from subleading corrections that are odd in $V_{12}$, which can cause the function to approach the asymptotic value from opposite directions on the two sides.
%
% \section{Eigenstate excitations as output}
% \label{app:eigenstate-output}
%

%
\bibliographystyle{apsrev4-2}
%\bibliography{refs}
\input{main_v3.bbl}

\end{document}